\documentclass[sigconf,screen,nonacm]{acmart}

\usepackage{booktabs}
\usepackage{graphicx}
\usepackage{subcaption}
\usepackage{xspace}
\usepackage{enumitem}
\usepackage{url}

\newcommand{\edkii}{EDK~II\xspace}

\begin{document}

\title{Enhancing Bug Report Templates in the TianoCore UEFI Firmware Development Community}

\author{Laura Baird}
\email{lbaird@uccs.edu}
\affiliation{%
  \institution{Department of Computer Science, University of Colorado Colorado Springs}
  \state{Colorado}
  \country{USA}
}

\author{Neelesh Reddybattula}
\email{nreddyba@uccs.edu}
\affiliation{%
  \institution{Department of Computer Science, University of Colorado Colorado Springs}
  \state{Colorado}
  \country{USA}
}

\author{Nazanin Siavash}
\email{nsiavash@uccs.edu}
\affiliation{%
  \institution{Department of Computer Science, University of Colorado Colorado Springs}
  \state{Colorado}
  \country{USA}
}

\author{Terrance E. Boult}
\email{tboult@uccs.edu}
\affiliation{%
  \institution{Department of Computer Science, University of Colorado Colorado Springs}
  \state{Colorado}
  \country{USA}
}

\author{Armin Moin}
\email{moin@purdue.edu}
\affiliation{%
  \institution{School of Applied and Creative Computing, Purdue University}
  \state{Indiana}
  \country{USA}
}

\renewcommand{\shortauthors}{Baird et al.}
\renewcommand{\shorttitle}{Enhancing Bug Report Templates}




\begin{abstract}
We propose enhancing the bug report templates in the GitHub Issues issue tracking system used by the TianoCore open-source community with the aim of improving the bug triage and resolution process. We analyze the bug repository data and find patterns of information that are useful for bug triage and fixing. However, some of them are only occasionally included in the free-form text of bug reports. Therefore, we propose adding a few new fields to the existing TianoCore bug report template. In this study, we focus on the key TianoCore project, \edkii, which constitutes the core of the UEFI firmware across various firmware vendors and original equipment manufacturers. This study is currently a work-in-progress. So far, we have interviewed a few developers to obtain their feedback and adjust the proposed approach. We are planning more interviews with the TianoCore community to conduct A/B tests and validate our approach to achieve effective and efficient bug triage and resolution.
\end{abstract}

\maketitle

\section{Introduction}
\label{sec:introduction}

Prior work has shown that developers value detailed reproduction steps, stack traces, test cases, and other structured evidence in bug reports in issue tracking systems. However, issue reporters often omit or bury this information in free-form text~\cite{Zimmermann+2010,Soltani+2020,Chaparro+2017,Chaparro+2019}. This mismatch is especially costly in firmware projects, where a report may depend on pre-operating-system execution state, firmware build options, CPU architecture, boot path, virtualized versus physical hardware, security-sensitive update, and runtime-service behavior. In such settings, the same symptom can imply different triage paths depending on whether it is a regression, whether it is architecture-specific, and whether the affected package is shared by downstream firmware stacks.

The TianoCore \edkii is an open-source reference implementation of the Unified Extensible Firmware Interface (UEFI). It is used across many firmware vendor stacks, which makes it a crucial component of the broader UEFI firmware supply chain. We focus on improving security and software maintenance workflows in this ecosystem. We make bug reports better structured to support bug triage and resolution. The \edkii project has migrated from Bugzilla to GitHub Issues. Its current bug report template has various structured fields, such as impacted packages, build targets, urgency, and maintainer feedback requests~\cite{TianoCore2026a,TianoCore2026b}. These fields are useful. However, our manual review and analysis of issues suggests that two firmware-specific signals remain under-structured: i) Bug report category, and ii) Affected CPU architectures. Both items matter for bug triage and resolution.

The contribution of this paper is threefold. First, it proposes to add new, pre-defined fields to bug reports in the \edkii issue tracking system to guide bug reporters in creating more complete issues. This way, we expect an increase in the effectiveness and efficiency of the work of bug triagers and bug fixers. Second, it proposes employing Artificial Intelligence (AI), in particular, Generative Pre-trained Transformer (GPT) Large Language Models (LLMs), to assist human triagers in their work by showing them some generated advisory comments. Last but not least, it presents a web-based prototype that enables large-scale validation and evaluation using feedback from the TianoCore developers community.

The remainder of this paper is structured as follows. Section \ref{sec:related-work} reviews related work. Furthermore, Section \ref{sec:approach} proposes the novel approach. We also present some preliminary validation results in Section \ref{sec:results}. Finally, Section \ref{sec:conclusion-future-work} concludes the paper and suggests future work.

\section{Related Work}
\label{sec:related-work}

Zimmermann et al.~\cite{Zimmermann+2010} showed that developers and bug reporters often disagree about what information is most useful and easiest to provide in bug reports. Soltani et al.~\cite{Soltani+2020} further studied the significance of bug report elements and found that elements such as reproduction steps and stack traces affect debugging and resolution outcomes. Chaparro et al.~\cite{Chaparro+2017,Chaparro+2019} developed techniques for detecting missing information in bug descriptions and assessing the quality of steps to reproduce. These studies have motivated our approach's premise that bug reporting interfaces should ask explicitly for information developers need rather than relying on reporters to determine what should be included.

GitHub issue templates and forms have become common mechanisms for improving report quality. Li et al.~\cite{Li+2023a} studied bug report templates on GitHub and found that templates can improve report comprehensibility and consistency. In a separate study, Li et al.~\cite{Li+2023b} (other authors despite the name similarity)  studied issue and pull-request templates more broadly, showing that template use has nuanced effects on issue discussion and resolution processes. Moreover, Zhang et al.~\cite{Zhang+2024} analyzed bug report templates across open-source software projects and derived design guidelines for maintaining them. Nikeghbal et al.~\cite{Nikeghbal+2024} explored automated generation of issue report templates. Song et al.~\cite{Song+2022} proposed interactive bug reporting through a chatbot that guides end users and verifies report quality. Our proposed approach differs from these general-purpose studies by focusing on firmware-specific triage fields and the TianoCore software supply chain context.

We assist in bug triage by providing human triagers with useful information. The bug triage process typically includes invalid and duplicate detection, assignee recommendation, and severity prediction. There exist prior work in the area of Mining Software Repositories (MSR) concerning semi-automated and/or automated bug triage. Qian et al.~\cite{Qian+2023} surveyed bug de-duplication methods. He et al. ~\cite{He+2020} proposed a convolutional neural network approach to duplicate bug report detection. Choetkiertikul et al.~\cite{Choetkiertikul+2021} recommended components for issue reports using deep learning. Some other recent LLM-based work, for example, by Acharya and Ginde ~\cite{AcharyaGinde2025} investigated generating or improving structured bug reports.

\section{Proposed Approach}
\label{sec:approach}

\begin{figure*}[t]
    \centering
    \includegraphics[width=0.72\linewidth]{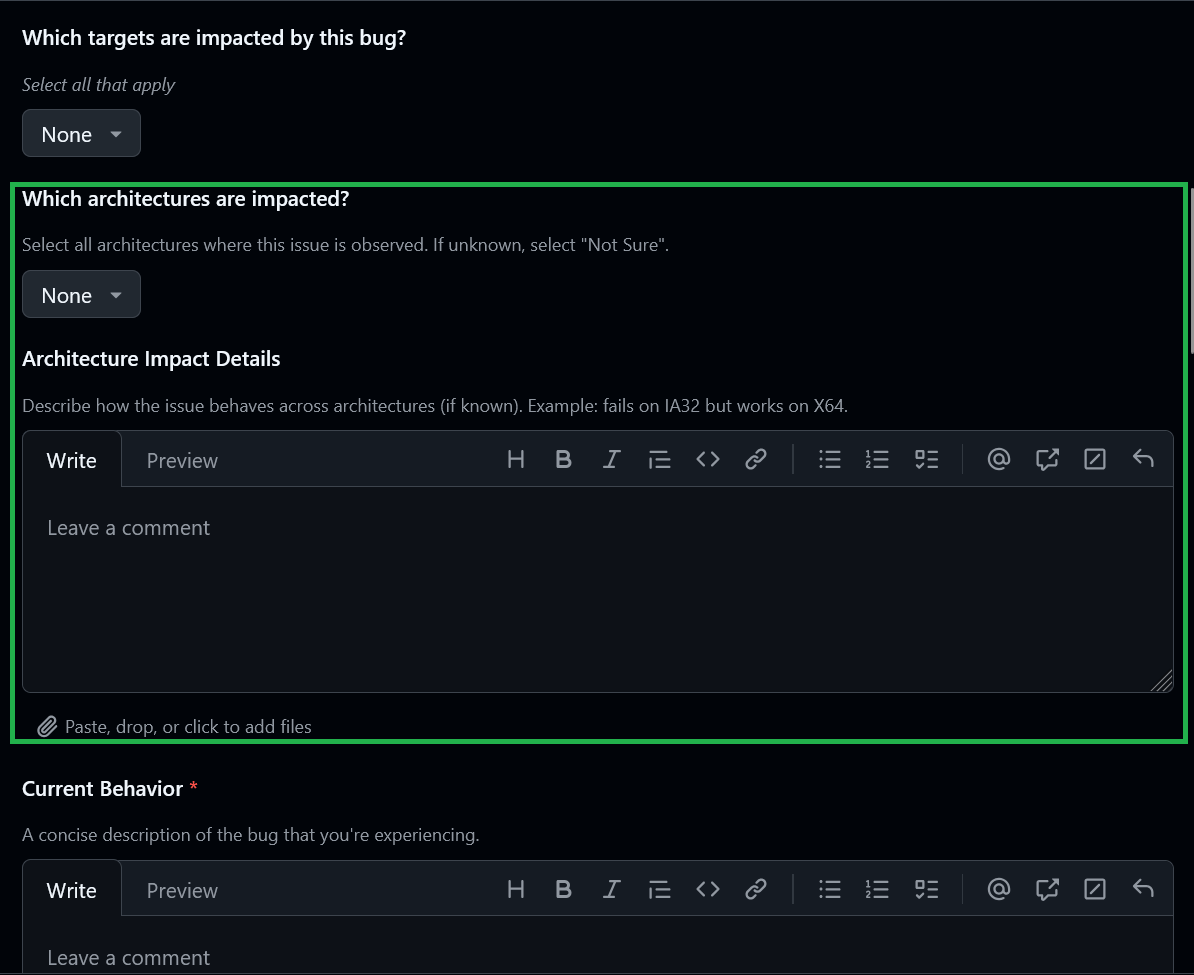}
    \caption{The impacted architectures added to the bug report template}
    \Description{A bug-report form containing selectable CPU architecture options and a text field asking how the reported behavior differs across architectures.}
    \label{fig:new-temp-2}
\end{figure*}

Based on repository analysis and developer feedback, we enable a number of domain-specific diagnostic fields in the TianoCore bug report structure. For instance, as depicted in Figure \ref{fig:new-temp-2}, we add a new field for the impacted architectures, with pre-defined options (multiple items can be selected). The options are the supported CPU architectures in TianoCore for the UEFI firmware. Here are the current options: IA32, X64, AARCH64, LOONGARCH64, ARM, RISCV64, and \emph{Not sure}. Besides that, a free-form textual field asks how behavior may differ across architectures. This captures qualitative information that cannot be expressed by a fixed dropdown list.

This can help bug triagers and bug fixers quickly determine the affected architectures. Moreover, it helps distinguish architecture-specific failures from general bugs. It may also offer benefits in the software supply chain security and management since companies in the ecosystem can easily decide whether an issue would be relevant to their repository.

\begin{figure*}[t]
    \centering
    \includegraphics[width=0.72\linewidth]{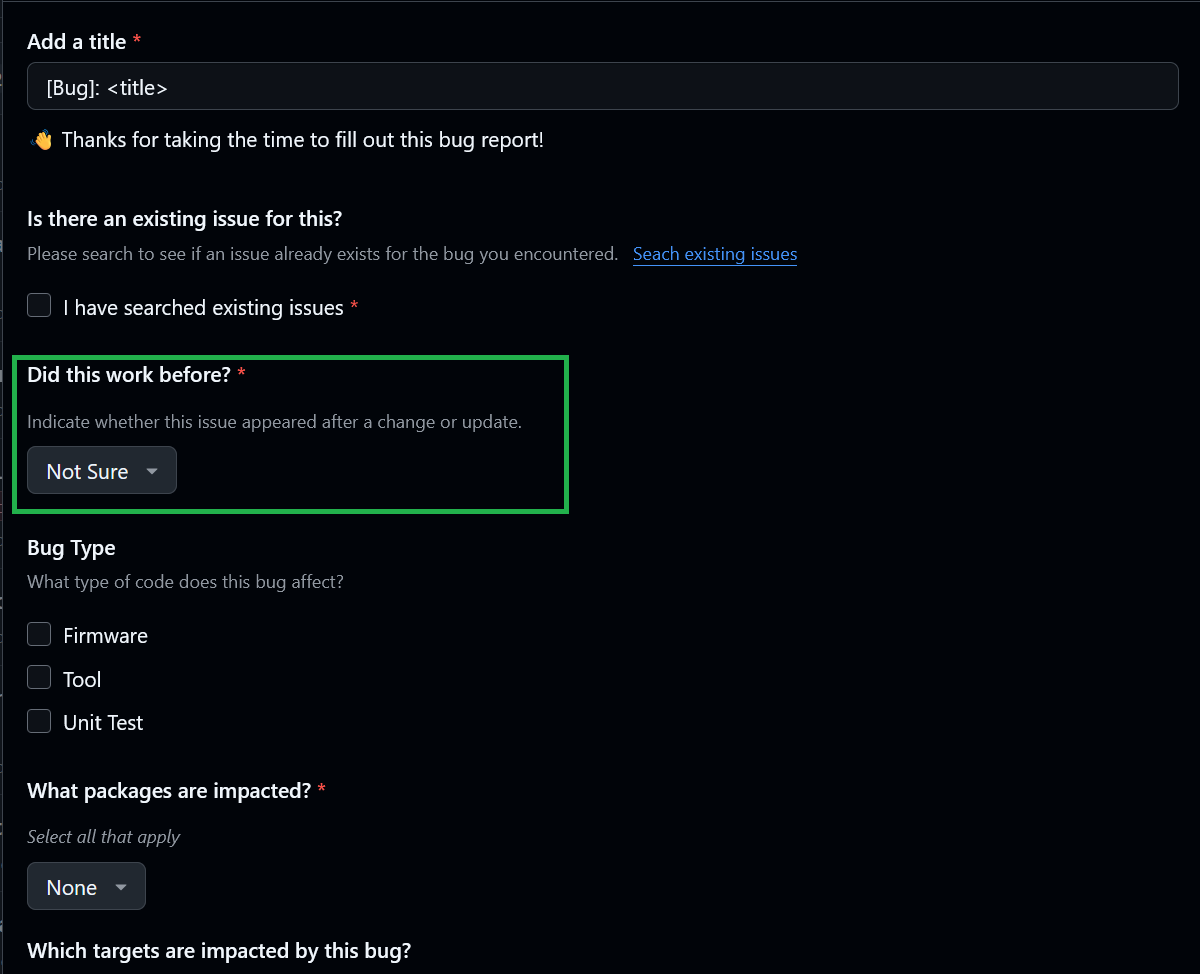}
    \caption{Identifying regression bugs in the new bug reports template}
    \Description{A bug-report form asking whether the behavior previously worked and requesting details about the last known working version or commit.}
    \label{fig:new-temp-1}
\end{figure*}

Additionally, we propose a new field in the bug reports that indicates the issue category. Issues in the issue tracking systems are commonly known as bug reports although not all of them are software defects (i.e., actual bugs). Some of them are descriptions of new-feature requests. Others may be issues that deserve special treatments. For example, TianoCore treats security vulnerabilities in a special manner and through a separate process different from non-security issues. Furthermore, we noticed that there might be some interest in labeling regression bugs. These are cases in which a software unit would work previously and would now be broken as a result of some development or maintenance activity. Therefore, we propose four possible options for the bug report category that can be in the form of a tag (a report may have several tags simultaneously): (i) New-feature requests, (ii) Functionality-related bugs, (iii) Regression bugs, and (iv) Security-related bugs. Note that security vulnerabilities are often handled through a separate process. The TianoCore community prefers the bug reporters to report such issues in a separate dedicated tab in the GitHub repository, named \textit{Security and quality}. There exists a documented process for responsible disclosure with an embargo period. In the current implementation of our prototype, only the regression category has been addressed. This is shown in Figure \ref{fig:new-temp-1}.

We also assist in bug triage without excluding human triagers in making decisions. This is achieved by advisory comments that are generated by AI-based methods. These comments automatically populate in the bug reports to provide additional information. However, they can be distinguished from the original bug report content. This information should be reviewed by developers (i.e., bug triagers, package maintainers, and bug fixers). We continuously seek feedback from them and loop this back into our AI-based method for improvements. In the present prototype, the AI-generated comments include suggestions from the AI method concerning the following: i) The validity of the bug report (is this really a bug?); ii) Duplicate detection (is there another issue in the repository reporting the same matter?); iii) Priority determination (is this a low/medium/high priority issue?); iv) Bug report assignment (who should fix this bug?). Furthermore, specifying the impacted packages and/or components automatically using our AI-based methods is a work-in-progress. The advisory outputs used in our A/B study were generated by TianoForge, an implemented pipeline contributed by co-author Nazanin Siavash~\cite{Siavash+2026}.

In addition, we provide a confidence score for the comments and determinations of the AI-based method. This confidence score is provided prior to the human user's (triager's, fixer's, or maintainer's) review. However, the review and feedback from the user will help adjust and enhance this feature in the future.

Moreover, our pilot interviews with developers indicated that they preferred explainable AI-based recommendations. Therefore, we offer some evidence and justification for each AI-generated advisory comment. For example, the generated text includes reasons for why the \textit{duplicate} or \textit{invalid} tag was suggested for an issue, which piece of bug report information supports a particular priority level for a bug report, or what historical data in the software repositories signal the fitness of an assignee to resolve the issue.

\section{Preliminary Results}
\label{sec:results}

In this section, we present some preliminary outcomes of this work-in-progress study.

\subsection{Methodology}

The key underlying research question is how we can adapt the current bug report structure in TianoCore to make bug triage and resolution more effective and efficient. The motivating factor for this was our observation concerning the very large number of open (and untriaged) issues in the issue tracking system, many of them high-priority bug reports.

To answer the above research question, we first review the information in the TianoCore software repositories, primarily the bug repository. We manually analyze 92 bug reports from the \edkii repository, drawn from a documented snapshot of all closed, \texttt{type:bug}-labeled GitHub issues that were originally opened on GitHub. We call the issues that were originally opened on GitHub Issues, \textit{GitHub-native}. The TianoCore community has migrated from the Bugzilla issue tracking system to GitHub Issues. Therefore, we call the issues that have been transferred from Bugzilla to GitHub Issues non-native to GitHub. The latter category is identifiable by a "Bugzilla Bug \#" reference in the issue title. The snapshot was captured on June 30, 2026, via the GitHub GraphQL API. The fetch script is available in the public source artifact~\cite{TianoShield2026}, and the raw issue data and coded results are archived in Harvard Dataverse~\cite{Baird+2026}. For each issue, we record whether the report describes a regression, whether it mentions CPU architecture information, and how architecture information relates to diagnosis. The corpus analysis is used to motivate the new template fields that can affect bug triage and resolution. Manual analysis was the primary signal for these labels; a separate LLM-assisted analysis was used only as an additional consistency check.

We also plan to conduct interviews, including A/B testing, with developers who engage in bug triage and/or bug fixing in TianoCore. Four pilot interviews have already been conducted to gain insight and prepare for the actual ones. We deploy a within-subjects A/B comparison. Participants review pairs of real \edkii issues in two representations shown under the blinded labels Template~1 and Template~2. The website deterministically randomizes both issue order and the mapping of the baseline and modified representations to these labels using each participant ID. The two issue examples in the current comparison tool are issue~11374, a runtime-driver memory-permissions regression, and issue~11882, a RedfishPkg IA32 NOOPT linker failure. We ask participants to think aloud, complete the on-screen questions, and answer verbal follow-up open questions about general preference, usefulness of the information, missing or confusing information, structure of the fields, information representation, etc. In the current pilot interviews, the population includes four participants enrolled in Ph.D. studies in a relevant field with one among them being a TianoCore security expert with industry development experience in TianoCore.

\subsection{Repository Analysis}

\begin{figure*}[t]
    \centering
    \includegraphics[width=0.75\linewidth]{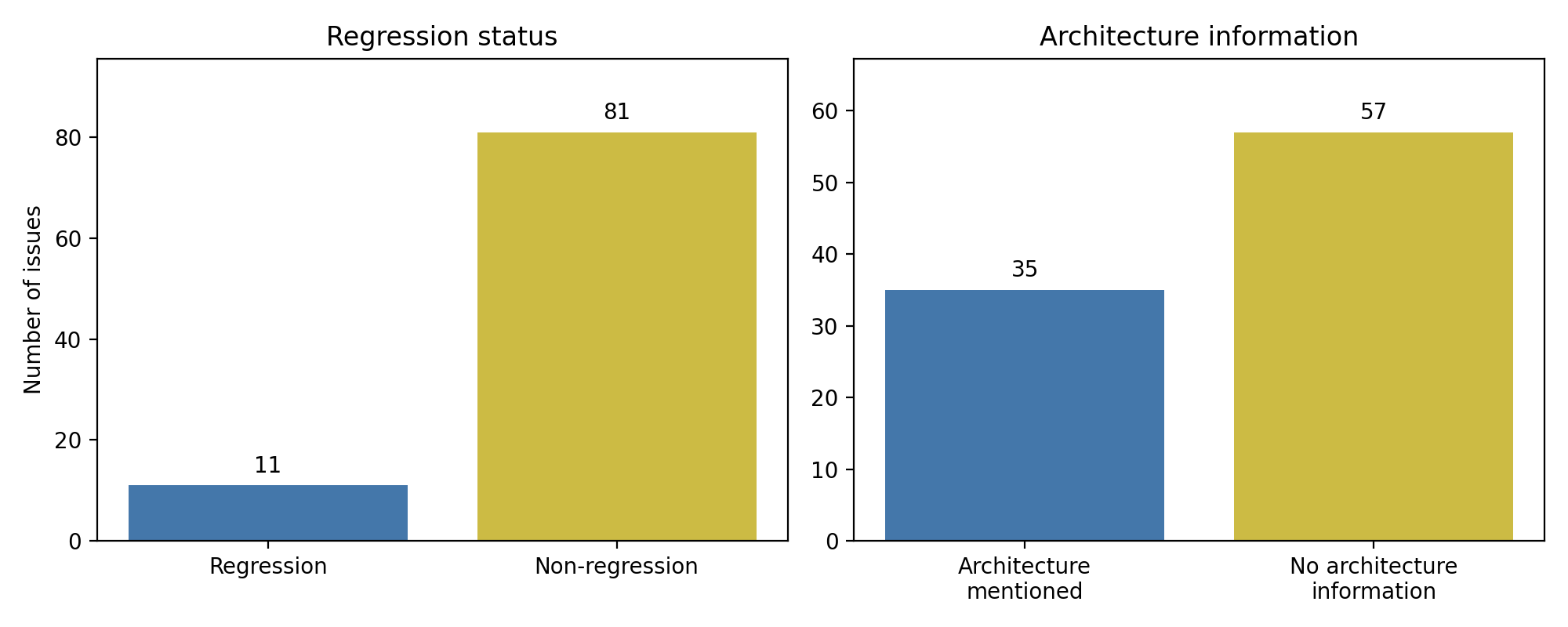}
    \caption{According to our bug repository analysis, regression and architecture information appear frequently enough to justify explicit fields for them in the bug reports.}
    \Description{Two bar charts. The first shows 11 regression and 81 non-regression issues. The second shows 35 issues with architecture information mentioned and 57 without.}
    \label{fig:issue-analysis}
\end{figure*}

Figure \ref{fig:issue-analysis} illustrates the frequency of regression bugs and the frequency of bug reports with architecture-related information in our selected corpus of 92 bug reports. 11 out of 92 issues were labeled as regressions. Also, 35 out of 92 bug reports mentioned architecture information. Furthermore, 16 issues used architecture information in a way that was trigger-specific or root-cause-specific. This indicates that architecture is not merely background context. Instead, in many cases it changes how the issue should be diagnosed.

\subsection{Pilot Interviews}

\begin{figure*}[t]
    \centering
    \includegraphics[width=0.88\linewidth]{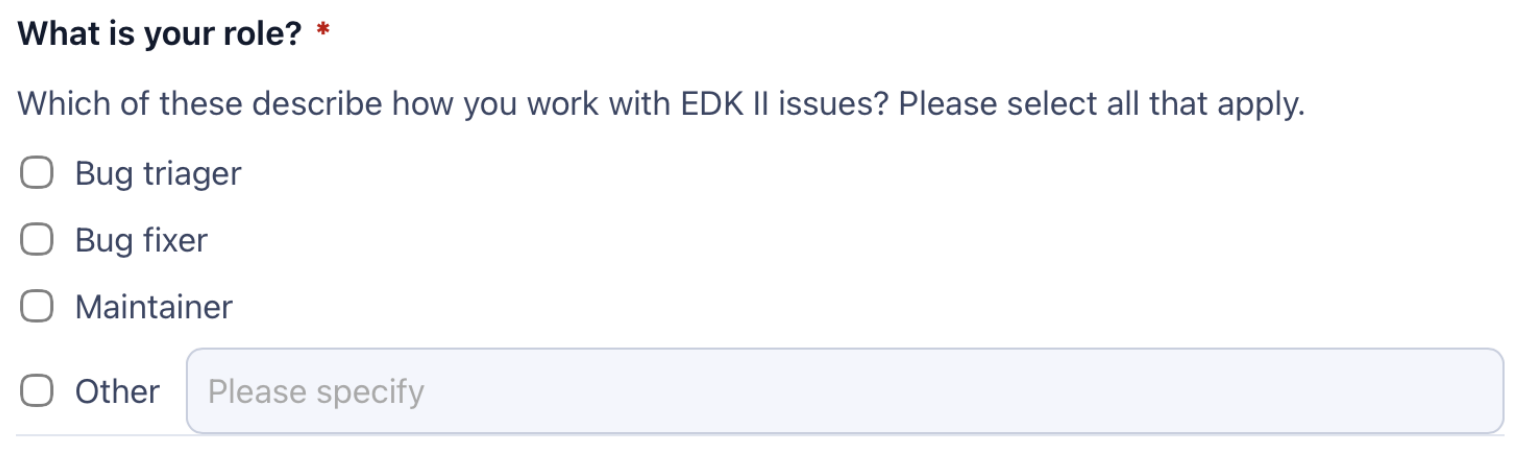}
    \caption{Screenshot from our web-based prototype for validation. Here, we are asking about the participant's primary role.}
    \Description{A study page asking the participant to identify their primary software-development, bug-reporting, triage, maintenance, or research role.}
    \label{fig:val-prot-p1}
\end{figure*}

\begin{figure*}[t]
    \centering
    \includegraphics[width=0.95\linewidth]{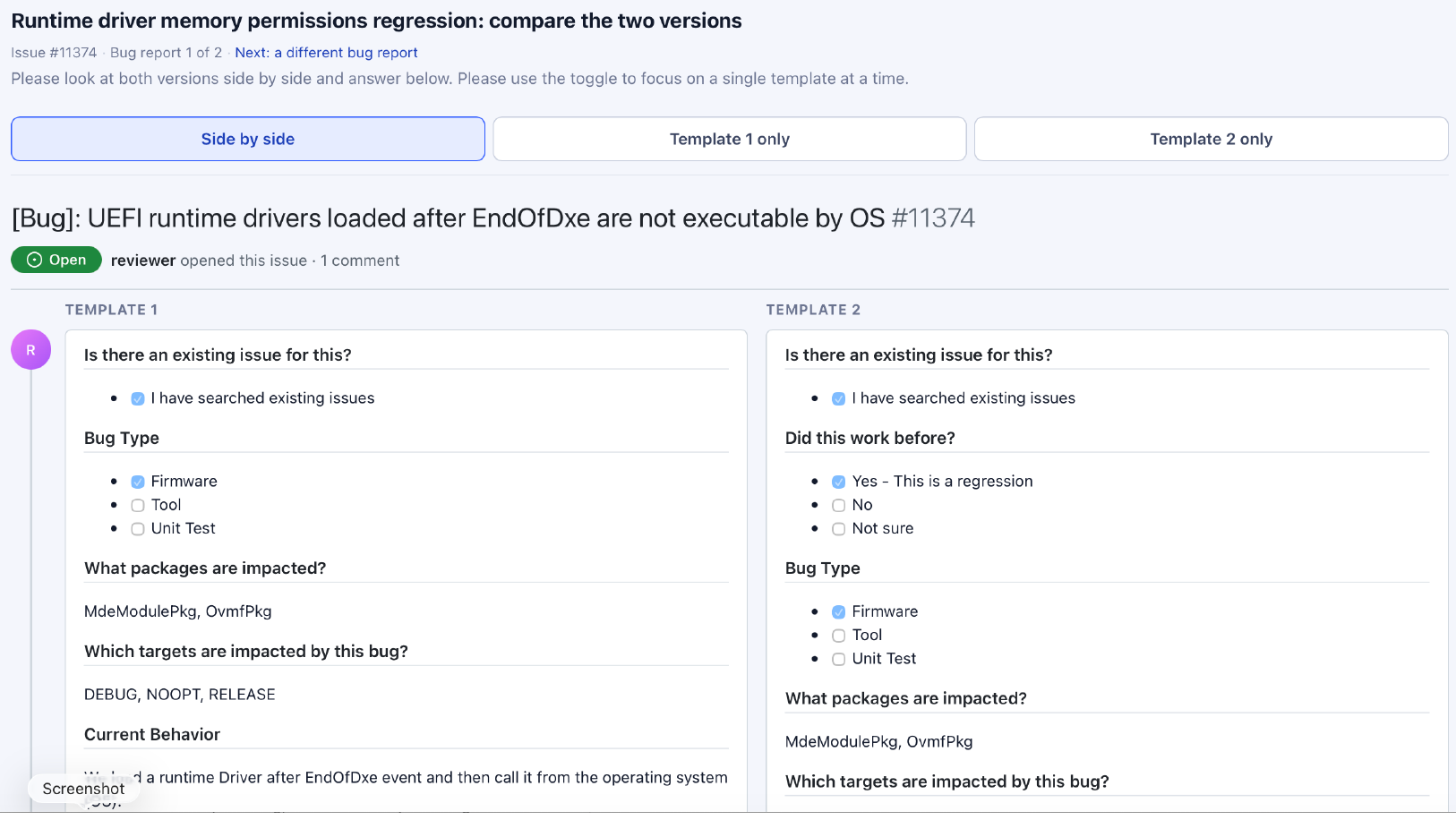}
    \caption{Screenshot from our web-based prototype for validation. The figure depicts two alternative views (original and enhanced with augmented fields) for the same bug report side by side.}
    \Description{A side-by-side comparison of two blinded representations of the same bug report, one using the existing structure and one containing additional structured triage information.}
    \label{fig:val-prot-p2}
\end{figure*}

During the interview sessions conducted through video calls over the Internet, we asked the participant to share their screen, open the web-based prototype shown in Figures \ref{fig:val-prot-p1} and \ref{fig:val-prot-p2} in their web browsers, read the provided information there, and answer the questions. Three out of four participants preferred the modified version with additional fields and AI support. However, one participant, who was the more experienced one, was critical of most additions and preferred the more concise version for bug triage. This participant was also critical of the original template and did not believe either of the variants would help them triage and/or resolve issues more effectively or efficiently. We intend to hold more discussions and feedback rounds with this participant, as well as the other TianoCore developers to incorporate their feedback into the prototypes.

\section{Conclusion and Future Work}
\label{sec:conclusion-future-work}

In this paper, we proposed an improved structure for bug report templates of the TianoCore \edkii that makes firmware-specific diagnostic fields explicit and integrates advisory AI comments for bug triagers and fixers. A manual analysis of 92 \edkii issues shows that regression and architecture information appear often enough to justify structured fields, while the exploratory pilot interviews provided formative design feedback on the new structured representation. 

This is a work-in-progress. We plan to conduct the main interviews with developers of TianoCore at a larger scale using the demonstrated web-based prototype. In addition, we will validate the quality of the AI-generated advisory comments with developers in various roles, including bug triagers, bug fixers, and package maintainers. Future iterations will place each recommendation and its linked supporting evidence before model-identification details to better support rapid issue scanning.

\section*{Software and Data Availability}
Our source code is publicly available at~\cite{TianoShield2026}; the corresponding replication data are archived in Harvard Dataverse~\cite{Baird+2026}.

\begin{acks}
This material is based upon work supported by the U.S. National Science Foundation (NSF) under Grant No. 2534021. Any opinions, findings, conclusions, or recommendations expressed in this material are those of the authors and do not necessarily reflect the views of the NSF. In preparing this work, we used generative AI models and tools, including OpenAI GPT and Anthropic Claude models, to assist in generating and revising code and text.

This work is accepted to be presented in the FTA 2026 workshop but is not published in the proceedings, according to the ACM SIGSOFT policy (\url{https://www2.sigsoft.org/policies/pcpolicy/}) that does not allow the work of organizers to be published in the workshop proceedings.
\end{acks}

\clearpage
\newpage
\bibliographystyle{ACM-Reference-Format}
\bibliography{references}

\end{document}